\documentstyle[emulateapj,onecolfloat,psfig]{article}   

\newcommand{\da}{d_A}

\newcommand{\vecx}{{\bf x}}
\newcommand{\veck}{{\bf k}}
\newcommand{\vecn}{{\bf n}}
\newcommand{\vecl}{{\bf l}}

\newlength{\tskip}
\newlength{\colwidth}

\newcommand{\beq}{\begin{equation}}
\newcommand{\eeq}{\end{equation}}
\newcommand{\beqa}{\begin{eqnarray}}
\newcommand{\eeqa}{\end{eqnarray}}

\def\simgt{\gtrsim}

\begin{document}
\twocolumn[   

\title{Second Order Corrections to Weak Lensing by Large-Scale Structure}
\author{Asantha Cooray$^{1,}$\altaffilmark{2} and Wayne Hu$^3$}
\affil{
$^1$Division of Physics, Mathematics and Astronomy, California
Institute of Technology, Pasadena, CA 91125.\\
$^3$Center for Cosmological Physics, 
Department of Astronomy and Astrophysics, and Enrico Fermi Institute, 
University of Chicago,
Chicago, IL 60637.\\
E-mail: asante@caltech.edu,whu@background.uchicago.edu}

\begin{abstract}
We calculate corrections to the power spectrum predictions of weak lensing by 
large scale structure due to higher order effects in the 
gravitational potential.
Using a perturbative approach to third 
order in transverse displacements, we calculate a second order correction to 
the angular power spectra of $E$ and $B$ mode shear and convergence  
resulting from dropping the so-called Born approximation, where
one integrates along the unperturbed photon path.
We also consider a correction to the power spectra from the coupling between
lenses at different redshifts.  Both effects generate $B$-mode
shear and the latter also causes a net rotation of the background galaxy
images. We show all these corrections are at least two orders of
magnitude below the convergence or $E$-mode power and hence relevant only
to future ultra high precision measurements.
These analytical calculations are
consistent with previous numerical estimates and
validate the use of current large scale structure weak lensing  predictions
for cosmological studies and future use of $B$-modes as a monitor of
systematic effects.
\end{abstract}

]

\altaffiltext{2}{Sherman Fairchild Senior Research Fellow} 

\section{Introduction}

As experiments that measure the distortion induced in distant
galaxy images by weak gravitational lensing from 
the large-scale structure of the universe
(e.g., \cite{Blaetal91} 1991; \cite{Mir91} 1991;
\cite{Kai92} 1992)  move from the current discovery phase
(\cite{Bacetal00} 2000; 
 \cite{Kaietal00} 2000;
 \cite{Witetal00} 2000;
 \cite{Vanetal00} 2000) 
into the precision measurement phase, it will become increasingly
important to quantify and separate subtle cosmological, astrophysical and 
instrumental effects that alter the statistics of the lensing observables.  

%

Current predictions of weak lensing statistics and
their utility in measuring fundamental cosmological parameters 
are based on several assumptions. 
In this Letter, we discuss two of  these assumptions and the
extent of their validity for calculations of the power spectra of
the shear and convergence. 
In the so-called Born approximation, one integrates the lensing
distortion over the unperturbed photon paths.
We relax this assumption using the transverse
deflection as a perturbative parameter. 
We show that the
correction to the Born approximation results in a generation of curl,
or magnetic-like, $B$-modes in the weak lensing shear.  Absent in 
first order lensing contributions, $B$-modes
are often used as a monitor of subtle systematic errors in the data.
Indeed, data from the current generation of surveys routinely show
$B$-modes in the shear field (e.g. \cite{Penetal02} 2002).
It will be important for future surveys to know the level 
at which lensing itself produces $B$-modes. 

An additional assumption is the neglect of the coupling between
lenses at two different redshifts. 
Unlike the correction to the Born
approximation, the lens-lens coupling results in both $B$-mode generation
and a net rotation of galaxy images. 
Since the rotational effect has been measured
in numerical simulations (\cite{Jaietal00} 2000), it provides a check on
the validity of our perturbative technique.   

The physical mechanism behind the two corrections discussed in 
the present paper were first considered by \cite{Beretal97} (1997) and \cite{Schetal98}
(1998). In both these studies, these two second order corrections were discussed 
as a possible contribution to the three point statistics of weak
lensing. The correction to the
weak lensing convergence skewness, however, is below a few percent.
In this paper, we concentrate on the two point statistics which requires
a third order expansion in the potential to complete a second order
expansion in the power.  

For illustrative purposes, we calculate these higher order effects for
a flat $\Lambda$CDM cosmology throughout,
with parameters $\Omega_c = 0.3$, $\Omega_b=0.05$, 
$\Omega_\Lambda=0.65$, $h=0.65$, $n=1$,
$\delta_H=4.2\times 10^{-5}$ and no tensor contribution.

\section{Statistical Properties}

In this section, we define the statistical properties of 
the lensing observables in angular Fourier space.  
In general, we can write the weak lensing deformation matrix that maps
between the source (S) and image  (I) planes, $\delta x_i^{\rm S} = A_{ij}
\delta x_j^{\rm I}$, as
\begin{eqnarray}
\label{A}
{\bf A} =
\left(
\begin{array}{cc}
1-\kappa-\gamma_1 & -\gamma_2-\omega\\
-\gamma_2+\omega & 1-\kappa+\gamma_1
\end{array}
\right) \, ,
\end{eqnarray}
where all components are functions of position on the sky ${\bf n}$.
Here, $\kappa$ is the
convergence and $\gamma_i$ are the two shear components
and $\omega$, the antisymmetric component induces a rotation in the
images
(e.g., \cite{BarSch00} 2000 and \cite{Mel99} 1999 for recent reviews).
The deformation component may be
isolated as 
\begin{equation}
\psi_{ab}(\vecn) = \delta_{ab} -A_{ab}(\vecn) \,.
\end{equation}
Following \cite{Kai98} (1998), we define the Fourier representation
of the deformation under the flat-sky approximation
\begin{equation}
\psi_{ab}(\vecl) = \int d{\vecn} e^{- i \vecl \cdot \vecn} \psi_{ab}(\vecn)\,.
\end{equation}
Statistical homogeneity requires that its two-point correlation satisfy
\begin{equation}
\left< \psi_{ab}^*(\vecl) \psi_{cd}(\vecl') \right>
= (2\pi)^2 \delta(\vecl-\vecl') C_{abcd}(\vecl)\,.
\end{equation}

The various components of $C_{abcd}$ define the two point 
statistical properties of the lensing observables.  
It is useful first to reexpress the shear components of 
$\psi_{ab}(\vecl)$ in terms of the coordinate-free 
$E$ and $B$ mode representation
(see, e.g. \cite{Ste96} 1996; \cite{Kametal98} 1998; \cite{HuWhi01} 2001)
\begin{eqnarray}
\epsilon(\vecl) &=& \cos 2 \phi_l \gamma_1(\vecl) +
\sin 2 \phi_l \gamma_2(\vecl) \, ,\nonumber \\
\beta(\vecl) &=& \cos 2 \phi_l \gamma_2(\vecl) -
\sin 2 \phi_l \gamma_1(\vecl)  \, .
\end{eqnarray}
where $\cos \phi_l = \vecl \cdot \hat{\bf x}_1$ where
$\hat{\bf x}_1$ is one of the orthogonal directions
on the sky. 


The power and cross spectra of the observable fields
$\alpha$, $\beta$ $=\kappa$, $\epsilon$, $\beta$, $\omega$ are defined as
\begin{eqnarray}
\left< \alpha^*(\vecl) \beta(\vecl') \right> \equiv
(2\pi)^2 \delta(\vecl-\vecl') C_l^{\alpha\beta}\,.
\end{eqnarray}
Since the power spectra are functions of the magnitude of $\vecl$ only,
one can take $\vecl \parallel \vecx_1$ 
to simplify calculations of the power spectrum
without loss of generality.
With this restriction, $\phi_l=0$ and 
we obtain for the power spectra
\begin{eqnarray}
C_l^{\kappa\kappa} 
 & = &{1 \over 4} [C_{1111}+2 C_{1122} + C_{2222}] \nonumber\,,\\
C_l^{\epsilon\epsilon} 
 &  =& {1 \over 4} [C_{1111}-2 C_{1122} + C_{2222}] \nonumber\,,\\
C_l^{\beta\beta} 
 &  =& {1 \over 4} [C_{1212}+2 C_{1221} + C_{2121}] \nonumber\,,\\
C_l^{\omega\omega} 
 &  =& {1 \over 4} [C_{1212}-2 C_{1221} + C_{2121}] \nonumber\,,
\end{eqnarray}
and similarly for the cross spectra. 
Cross spectra between $\kappa$ or $\epsilon$ with
$\beta$ or $\omega$ vanish if the statistical properties are invariant under
inversion of the coordinates.

\section{Calculational Technique}

It remains to evaluate the deformation power spectrum $C_{abcd}$
to second order in perturbation theory.
The deformation tensor for sources at a single redshift 
is given implicitly by (see e.g. \cite{Schetal98} 1998)
\begin{equation}
\psi_{ab}(\vecn,\chi_s) = 2 \int
d \chi g(\chi,\chi_s) 
\Phi_{,ac}(\vecx;\chi) [\delta_{cb} + \psi_{cb}(\vecn,\chi)] \,,
\label{eqn:psiij}
\end{equation} 
where commas represents spatial derivatives. 
Repeated indices are summed over the 2 transverse directions.
The presence of $\psi$ in the integral reflects a foreground lens
affecting the deformation from a more distant lens - or ``lens-lens coupling''.
The lensing efficiency is
\begin{eqnarray}
g(\chi',\chi)& = & \frac{d_A(\chi') d_A(\chi-\chi')}{d_A(\chi)}\,, \quad 
			\chi'<\chi \nonumber\,,\\
	      & = & 0\,, \quad \chi' \ge \chi\,.
\end{eqnarray}
Here the gravitational potential $\Phi$ is evaluated at a deflected 
position\footnote{Formally $d_A(\chi)\vecn \rightarrow d_A(\chi)\vecn_\perp
+ \chi \vecn_\parallel$ in an open universe but components parallel to 
the fiducial line of sight drop out in the Limber approximation.} 
\begin{eqnarray}
 \vecx( \vecn,\chi) &=& \vecn d_A(\chi) + \delta \vecx(\vecn,\chi) \,, \nonumber\\
 \delta x_a(\vecn,\chi) &= &- 2 \int d\chi' g(\chi',\chi)
  {d_A(\chi) \over d_A(\chi')} \Phi_{,a}(\vecx;\chi')\,,
\label{eqn:deflect}
\end{eqnarray}
where the deflections are confined to the transverse plane and
$\chi$ is the conformal distance or
lookback time, from the observer, given by
\begin{equation}
\chi(z) = \int_0^z {dz' \over H(z')} \,,
\end{equation}
and the analogous angular diameter distance
\begin{equation}
\da(\chi) = H_0^{-1} \Omega_K^{-1/2} \sinh (H_0 \Omega_K^{1/2} \chi)\,,
\end{equation}
with $\Omega_K = 1-\Omega_{\rm tot}$ as the space curvature parameter.  
We will occasionally suppress
the radial/temporal argument $\chi$ where no confusion will arise.

The familiar form of the lensing observables comes about by
keeping only first order terms in the potential.  In other words, they are
calculated under the so-called Born approximation where the potential
is evaluated on the undeflected path and with the neglect of lens-lens coupling.
Higher order corrections can then be calculated by iterative 
correction of ${\bf x}$ and $\psi_{ab}$ in Eq.~(\ref{eqn:psiij}).

We shall see that in general the deformation may be expressed 
as a line-of-sight projection of a source field
\begin{equation}
\psi_{ab}(\vecn) = 2 \int d\chi g(\chi,\chi_s) S_{ab}(\vecn d_A;\chi) \,,
\end{equation}
whose power spectrum may be evaluated under the Limber approximation 
(\cite{Kai92} 1992) from the power spectrum of the source field
\begin{eqnarray}
\left< S_{ab}^*(\vecl;\chi) S_{cd}(\vecl';\chi')\right>
&=& (2\pi)^2 \delta(\vecl-\vecl') \delta(\chi-\chi')\nonumber\\
&&\quad \times
d_A(\chi)^{-6} 
		P_{abcd}(\vecl;\chi) \,,
\end{eqnarray}
to be
\begin{equation}
C_{abcd}(\vecl) = 4 \int d\chi { g^2(\chi,\chi_s) \over d_A^{6}(\chi)} 
P_{abcd}(\vecl;\chi)\,.
\end{equation}

For the first order term $S_{ab}^{(1)} = \Phi_{,ab}$ and
\begin{equation}
P_{abcd}^{(11)} = l_a l_b l_c l_d P_{\Phi\Phi}\left({ l \over d_A}\right)\,,
\end{equation}
where $P_{\Phi\Phi}(k)$ is the spatial 
power spectrum of the potential fluctuations evaluated under the
non-linear scaling relations for the density power spectrum (\cite{PeaDod96} 1996).

As discussed in the previous section, one may evaluate the power spectra
of the observables under the constraint $l_1=l$ and $l_2=0$, from which
one immediately obtains
\begin{eqnarray}
C_l^{\kappa\kappa_{11}} =
l^4 \int_0^{\chi_s} d\chi \frac{g^2(\chi,\chi_s)}{\da(\chi)^6}
P_{\Phi\Phi}\left(\frac{l}{\da(\chi)};\chi\right) \, ,
\end{eqnarray}
and $C_l^{\epsilon\epsilon_{11}} = C_l^{\kappa\kappa_{11}}$
while $C_l^{\beta\beta_{11}}=C_l^{\omega\omega_{11}}=0$. Here
the subscripts $nn$ denote $n$th order term expansion in the potentials.
The lack of $B$-modes in the signal, to first order in the perturbations, 
has been considered as a possible test of instrumental and 
astrophysical systematic effects.   

In general if $S_{ab}$ is symmetric under interchange of $a$ and $b$,
power spectra involving the rotation $\omega$ 
vanish.  If $P_{abcd}$ is symmetric under
the exchange of all indices then $C_l^{\kappa\kappa} 
= C_l^{\epsilon\epsilon}+C_l^{\beta\beta}$. 
Finally if $P_{1122}=P_{1221}$ then $C_l^{\kappa\kappa}+C_l^{\omega\omega}
= C_l^{\epsilon\epsilon}+C_l^{\beta\beta}$.  These relationships are useful
for checking consistency in the higher order terms.

\subsection{Born approximation}

Corrections to the Born approximation can be calculated by Taylor expanding
the potential in Eq.~(\ref{eqn:psiij}) to second order in the deflection of
Eq.~(\ref{eqn:deflect})
\begin{eqnarray}
\Phi(\vecn d_A+\delta \vecx)& = &\Phi(\vecn d_A) + \delta x_a \Phi_{,a}(\vecn d_A)
\nonumber\\
&& \quad +\frac{1}{2} \delta x_{a} \delta x_{b} \Phi_{,ab}(\vecn d_A) + \ldots \, .
\end{eqnarray}
The last term is effectively third order in the potential but must be kept
since it can couple with the first order deformation in the power spectrum.
The corrections may be represented as second order and third order source
fields
\begin{eqnarray}
S_{ab}^{(2)}(\vecn) &=&- 2 \Phi_{,abc}(\vecn d_A) B_{c}(\vecn)\,,\nonumber\\
S_{ab}^{(3)}(\vecn) &=& 2 \Phi_{,abcd}(\vecn d_A) B_{c}(\vecn) B_d(\vecn)\,,
\label{eqn:bornsource}
\end{eqnarray}
with the Born correction 
\begin{equation}
B_c(\vecn;\chi) \equiv 
\int d \chi' g(\chi',\chi) { d_A(\chi) \over d_A(\chi')} 
\Phi_{,c}(\vecn d_A(\chi');\chi')\,.
\end{equation}

The second-second order terms in the power spectrum become
\begin{equation}
P_{abcd}^{(22)} = 4 \int {d^2 l' \over (2\pi)^2} 
l_{a}' l_{b}' l_{c}' l_{d}'  (\vecl' \cdot \vecl'')^2
M(l',l'') \, ,
\end{equation}
with $\vecl''= \vecl-\vecl'$.
The mode coupling integrand is given by
\begin{eqnarray}
M(l,l';\chi) &=& 
			\int d\chi' { g^2(\chi,\chi') \over d_A^{6}(\chi')}
			  P_{\Phi\Phi}\left({l' \over d_A(\chi')};\chi'\right)
			\nonumber\\
&&\quad \times
			P_{\Phi\Phi}\left({l \over d_A(\chi)};\chi\right)  \, .
\label{eqn:mode}
\end{eqnarray}
In terms of the lensing observables,
\begin{eqnarray}
C_l^{\alpha\beta_{22}} &=& 
4 \int d\chi \frac{g^2(\chi,\chi_s)}{\da(\chi)^6}
\int {d^2 l' \over (2\pi)^2} 
l'^4 G_\alpha^{(22)} G_\beta^{(22)}
\nonumber \\ &&\quad \times (\vecl' \cdot \vecl'')^2 M(l',l'') \, .
\end{eqnarray}
The geometric factors for the specific observables are
\begin{eqnarray}
G_\kappa^{(22)} &=& 1, \nonumber\\
G_\epsilon^{(22)} &=& \cos\phi_{l'}, \nonumber\\
G_\beta^{(22)} &=& \sin\phi_{l'}, \nonumber\\
G_\omega^{(22)} &=& 0, 
\label{eqn:borng}
\end{eqnarray}
where we have used the condition $\phi_l=0$.  
Notice that the parity violating cross terms vanish after 
integration over the azimuthal angle.

Likewise the first-third order source term becomes
\begin{equation}
P_{abcd}^{(13)} = -4 l_{a} l_{b} l_{c} l_{d} 
	\int {d^2 l' \over (2\pi)^2} (\vecl \cdot \vecl')^2 
	M(l,l')\,.
\end{equation}
Note that there is no mode coupling in the third order term. 
For the lensing observables,
\begin{eqnarray}
C_l^{\alpha\beta_{13}} 
&=& 
-4 \int d\chi \frac{g^2(\chi,\chi_s)}{\da(\chi)^6}
\int {d^2 l' \over (2\pi)^2} 
l^4 G_\alpha^{(13)} G_\beta^{(13)} \nonumber\\
&&\quad\times
(\vecl \cdot \vecl')^2 M(l,l')  \, ,
\end{eqnarray}
where
\begin{eqnarray}
G_\kappa^{(13)} &=& G_\epsilon^{(13)} =  1, \nonumber\\
G_\beta^{(13)} &=&  G_\omega^{(13)} = 0 \, .
\end{eqnarray}
Notice that contributions to the power from modes with $\vecl' 
= \vecl-\vecl'' \approx \vecl$ in Eq.~$(22)$ is cancelled by 
contributions from Eq.~$(13)$ for $\kappa\kappa$ and $\epsilon\epsilon$.  
Finite $\beta\beta$ power
is generated by the Born corrections but is also highly suppressed
geometrically.
There is also no contribution to the rotational power
from the second order Born approximation.
The rotational power spectrum as measured in simulations by \cite{Jaietal00}
(2000) should not be interpreted as a test of the Born approximation.

\begin{figure}[!t]
\centerline{\psfig{file=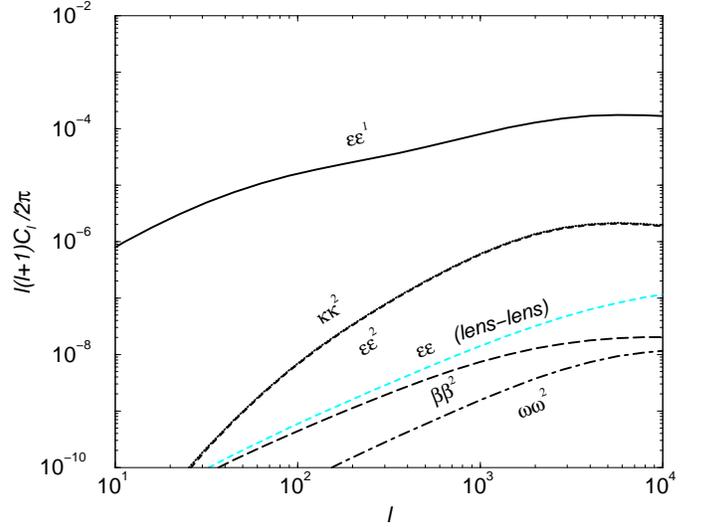,width=3.5in,angle=-90}}
\caption{Angular power spectra of weak lensing. The solid 
line is the well known first order result.
The dot, dashed, long-dashed and dot-dashed lines
are the total, Born approximation plus lens-lens coupling,
second order contribution to lensing observables involving
convergence, shear $\epsilon$-modes, shear $\beta$-modes, and
rotation, respectively. For comparison, the gray dashed line shows the
contribution to shear $\epsilon$-modes from lens-lens coupling only; $\kappa \kappa$ and $\beta \beta$ scale similarly.
Note that the rotational power spectrum, $\omega \omega$, is also
generated by lens-lens coupling only.}
\label{fig:cl}
\end{figure}

\subsection{Lens-Lens coupling}

Lens-lens coupling involves the iterative correction of the lensing
deformation of distant lenses due to the deformation from foreground
lenses in Eq.~(\ref{eqn:psiij}).  The second and third order 
corrections become
\begin{eqnarray}
S_{ab}^{(2)}(\vecn) &=& 
-\Phi_{,ac}(\vecn d_A;\chi) \nonumber\\
&&\times 2 \int d\chi' g(\chi',\chi)  
\Phi_{,cb}(\vecn d_A(\chi');\chi')\,,\nonumber\\
S_{ab}^{(3)}(\vecn) &=& 
\Phi_{,ac}(\vecn d_A;\chi) \nonumber\\
&&\times
2 \int d\chi' g(\chi',\chi)  
\Phi_{,cd}(\vecn d_A(\chi');\chi')
\nonumber\\
&&\times 
2 \int d\chi'' g(\chi'',\chi')  
\Phi_{,db}(\vecn d_A(\chi'');\chi'')\,.
\label{eqn:lenssource}
\end{eqnarray}

The second-second order terms in the power spectrum become
\begin{equation}
P_{abcd}^{(22)} = 4 \int {d^2 l' \over (2\pi)^2} 
l_{a}' l_{b}'' l_{c}' l_{d}''  (\vecl' \cdot \vecl'')^2
M(l',l'')\,,
\end{equation}
with $\vecl'' = \vecl -\vecl'$ and recall that the 
mode coupling integrand was defined in Eq.~(\ref{eqn:mode}).
The first-third order term vanishes under the Limber approximation
since lenses at the same redshift cannot lens each other.

The results for the power spectra are
\begin{eqnarray}
C_l^{\alpha\beta_{22}} &=& 
4 \int d\chi \frac{g^2(\chi,\chi_s)}{\da(\chi)^6}
\int {d^2 l' \over (2\pi)^2} 
l'^2 l''^2
G_\alpha^{(22)} G_\beta^{(22)}
\nonumber \\
&&\quad \times (\vecl' \cdot \vecl'')^2 M(l',l'')\,,
\end{eqnarray}
with the geometric factors
\begin{eqnarray}
G_\kappa^{(22)} &=& \cos(\phi_{l'} - \phi_{l''})\,, \nonumber\\
G_\epsilon^{(22)} &=& \cos(\phi_{l'} +\phi_{l''})\,, \nonumber\\
G_\beta^{(22)} &=&  \sin(\phi_{l'} +\phi_{l''})\,,  \nonumber\\
G_\omega^{(22)} &=& \sin(\phi_{l'} -\phi_{l''})\, .
\label{eqn:llg}
\end{eqnarray}
Recall we have set $\phi_{l}=0$. As derived, lens-lens coupling generates power in all the lensing observables. 
Again, notice that the parity violating terms oscillate leaving
negligible contribution after azimuthal integration.
%
\subsection{Born-Lens coupling}

There are also cross terms between corrections to the Born approximation and
lens-lens coupling.
The second-second order term follows directly from Eqs.~(\ref{eqn:bornsource}) 
and (\ref{eqn:lenssource})
\begin{equation}
P_{abcd}^{(22)} = 4 \int {d^2 l' \over (2\pi)^2} 
(l_{a}' l_{b}'  l_{c}' l_{d}''
+
 l_{a}' l_{b}'' l_{c}' l_{d}')
  (\vecl'\cdot \vecl'')^2
M(l',l'')\,,
\end{equation}
where again $\vecl'' = \vecl -\vecl'$.

The second-second order terms for the
lensing observables are likewise simply related to
the pure Born and lens-lens corrections 
\begin{eqnarray}
C_l^{\kappa\kappa_{22}} &=& 
4 \int d\chi \frac{g^2(\chi,\chi_s)}{\da(\chi)^6}
\int {d^2 l' \over (2\pi)^2} 
l'^3 l''
G_\alpha^{(22)} G_\beta^{(22)}
\nonumber\\
&&\quad \times (\vecl' \cdot \vecl'')^2 M(l',l'')  \, ,
\end{eqnarray}
where the geometric factors are simply products of the individual Born (B)
and lens-lens (LL) factors
\begin{eqnarray}
G_\alpha^{(22)} G_\beta^{(22)} =
G_\alpha^{(22)} \Big|_{\rm B}  G_\beta^{(22)} \Big|_{\rm LL} +
G_\alpha^{(22)} \Big|_{\rm LL}  G_\beta^{(22)} \Big|_{\rm B}  \, ,
\end{eqnarray}
given in Eqns.~(\ref{eqn:borng}) and (\ref{eqn:llg}).

For the third order term
\begin{eqnarray}
S_{ab}^{(3)}(\vecn) &=& 2 \Phi_{,acd}(\vecn d_A) B_d(\vecn) \nonumber\\
&&\times 2 \int d\chi' g(\chi',\chi)  
\Phi_{,cb}(\vecn d_A(\chi');\chi')\,,
\end{eqnarray} 
and 
\begin{equation}
P_{abcd}^{(13)} = 16 
	\int {d^2 l' \over (2\pi)^2} 
	(l_{a} l_{b} l_{c} l_{d}' +
	 l_{a} l_{b}'l_{c} l_{d}) 
	(\vecl \cdot \vecl')^2 
	M(l,l')\,.
\end{equation}
This implies that all power spectra of lensing observables
vanish after integration over the azimuthal angle $\phi_{l'}$.

\section{Results \& Discussion}

Here, we presented a 
general discussion of the accuracy of first order calculations of statistical
properties of weak lensing by large-scale structure. Using a perturbative approach to third
order in transverse displacements, we calculated the leading order
corrections to angular power spectra of the shear and convergence that
results from dropping the Born approximation.
We also considered the coupling between lenses at two different redshifts
in its contribution to the power spectra of lensing observables. 
Both effects generate power in the $B$-mode shear which places an
ultimate limit on its use as a monitor of systematics or for the
search for weak lensing by gravitational waves. 
In addition, lens-lens coupling results in  a net rotation of the background galaxy
images. 

In Fig.~\ref{fig:cl}, we summarize our results and show the total contribution to the
lensing observables at the second order level in power from corrections involving the Born approximation and lens-lens
couplings. As shown, these 
corrections are at least two orders of magnitude below the power in the convergence, 
or $E$-mode 
shear, at the first order level.  While these corrections are larger than 
cosmic variance at $\ell \simgt 100$, they are unlikely to affect the 
interpretation of the next generation of surveys. 
Our analytical calculations are consistent with previous numerical estimates:
the rotational power spectrum generated by the coupling between two lenses agree with numerical
measurements by \cite{Jaietal00} (2000). These results also
validate the use of current large scale structure weak lensing predictions for cosmological studies.
The relatively small contribution to the lensing observables suggest that higher order corrections 
related to the Born approximation and coupling between two lenses are
unlikely to affect the current estimates of weak lensing statistics
and place only very mild limitations on the use of $B$-modes as a 
monitor of systematics in future surveys.

\smallskip
{\it Acknowledgments:} WH was supported by
NASA NAG5-10840 and the DOE OJI program. AC was supported by the Sherman Fairchild foundation and the DOE.
We thank Marc Kamionkowski for useful discussions.

\end{document}